\documentclass[aps,prl,twocolumn,nofootinbib,superscriptaddress,amssymb]{revtex4}
\usepackage{color}
\usepackage{epsfig}
\usepackage{graphicx}
\usepackage{epstopdf}

\def\be{\begin{equation}}
\def\ee{\end{equation}}

\begin{document}

\title{Hypercharged Anomaly Mediation}

\author{Radovan Derm\' \i\v sek}

\affiliation{School of Natural Sciences, Institute for Advanced Study, Princeton,
NJ 08540}

\author{Herman Verlinde}
\author{Lian-Tao Wang}
\affiliation{Department of Physics, Princeton University, Princeton,
NJ 08544}

\date{November 19, 2007}

\begin{abstract}
We show that, in string models with the MSSM residing on D-branes, the 
bino mass can be generated in a geometrically separated hidden sector. 
Hypercharge mediation thus naturally teams up with anomaly mediation.
The mixed scenario predicts a distinctive yet viable superpartner 
spectrum, provided that the ratio $\alpha$ between the bino 
 and gravitino mass lies in the range $0.05\lesssim |\alpha|
\lesssim
0.25$ and $m_{3/2} \gtrsim 35$ TeV. 
We summarize some of the experimental signatures
of this scenario.
\end{abstract}

\pacs{}

\maketitle

\def\cAv{{A_{{}_V}\! }}
\def\cAh{{A}_{{}_H}}
\def\fv{{f}_{{}_V}\!}
\def\fh{{f}_{{}_H}\!}
\def\cW{{\mbox{\small $W$}}}

\def\cL{{\cal L}}
\def\MSSM{{\rm mssm}}
\def\HIDDEN{{\rm hidden}}
\def\ccc{{\mbox{\small $C$}}}
\def\wwwedge{\! \wedge}
\def\spc{}
\def\is{\!& \!=\! & \!}
\def\ba{\begin{eqnarray}}
\def\ea{\end{eqnarray}}





{\it Introduction:}
In supersymmetric models, the superpartner spectrum is dictated
by the mechanism by which supersymmetry (SUSY) breaking is transmitted
to the Standard Model.   Available scenarios fall into two main categories. 
In Planck scale mediation, SUSY is broken at a high scale and
transmitted to the visible sector via Planck scale modes. Alternatively, in gauge-mediation, SUSY breaking takes place at a lower scale and is communicated 
 via gauge theory degrees of freedom. 

An attractive geometric set-up for messenger mediated SUSY breaking is via
string models in which the visible and hidden sectors are both localized on branes \cite{Blumenhagen:2006ci}.
To realize gauge mediation, the hidden and visible branes must
be placed at a small relative distance $d \ll \ell_s$, 
so that the messengers arise as light open strings 
that stretch between the two. 
In Planck scale mediation, on the other hand, the hidden and visible sector
are typically taken to be separated by a distance $d >\ell_s$, and
SUSY breaking is transmitted via closed string modes.   Since 
the properties of the closed string messengers depend
sensitively on details of the Planck scale 
geometry, the SUSY flavor and CP problems -- the
strict bounds on  
flavor and CP violations from new physics~-- impose severe 
constraints on high scale mediation scenarios.

The most elegant Planck scale mediation mechanism is anomaly mediation
(AMSB) \cite{amsb}.
This scenario, in which the soft mass parameters are generated via the rescaling anomaly, has several attractive features: it  has just one free parameter (the gravitino mass $m_{3/2}$), avoids the flavor problem, and
the predicted spectrum is UV insensitive. 
The anomaly induced contributions are always present whenever
SUSY is broken; anomaly mediation refers to the case when these terms 
dominate the observable SUSY breaking effects. For this
to happen, the SUSY breaking scale needs to be high, while all
effects due to tree-level gravity mediation are suppressed.

It is non-trivial to find string scenarios where these conditions are
satisfied   \cite{Anisimov:2002az}.
The most promising  set-up is to localize the SUSY breaking 
at the bottom of a strongly warped hidden region,
geometrically separated from the visible region where the MSSM resides.
The warping effectively filters out all unwanted 
observable contributions due to tree-level gravity mediation 
\cite{Kachru:2007xp}. In the dual perspective, 
the warped throat describes a strongly coupled hidden CFT and the sequestering 
takes place due to RG suppression of the dangerous cross couplings
\cite{Schmaltz:2006qs}.

Recent studies have shown that this warped sequestering mechanism  
plausibly creates the pre-conditions for realizing anomaly 
mediation in string theory 
\cite{Kachru:2007xp}. 
This insight opens up interesting new
avenues for string model building. 
However, minimal AMSB predicts a negative mass squared
for the sleptons \cite{amsb}. Therefore, one needs to include at least
one other type of  
SUSY breaking effect.  In this note, we will identify an attractively 
simple, string motivated mediation mechanism, that naturally teams up
with anomaly mediation, 
and cures the tachyonic slepton problem.

\medskip

{\it Hypercharged Anomaly Mediation:}
Suppose that the MSSM is realized on a local stack of D-branes
\cite{Blumenhagen:2006ci}.
The closed string moduli that govern the MSSM couplings are then 
typically localized near the MSSM
branes.  The sequestering mechanism  
relies on this fact.  However, there 
is one geometrically well-motivated exception: the hypercharge 
gauge coupling may depend on moduli that are localized far from the
visible region.

Hypercharge $U(1)_Y$ is carried by a particular Dp-brane inside the MSSM stack. (One usually
considers D6-branes in IIA, and D5-branes in IIB.)
To ensure that $U(1)_Y$
survives as a low energy gauge symmetry, the hypercharge brane needs to
wrap a homologically trivial cycle \cite{Buican:2006sn}.
To arrange for this, one typically introduces 
a partner brane in the same homology class \cite{Blumenhagen:2006ci}, which could be part of a hidden sector. 
In this setup, depicted in Fig. 1, the two branes each 
produce their own $U(1)$ vector multiplet, $\cAv$ and ${\cAh}$, and the open string action
 splits up as (here $Q$ encodes all other MSSM fields) 
\be
\label{open}
\cL_{\MSSM}(Q, \cAv) + \cL_{\HIDDEN}(
{\cAh}). 
\ee
As explained in \cite{rrform},
the interaction with the closed string sector enforces a
low energy field identification between $\cAv$ and $\cAh$. 
This phenomenon is specific to $U(1)$ gauge fields. The mechanism
relies on the CS coupling $\int  C_{p-1}  \wedge
{\rm tr} F$. 
Here $C_{p-1}$ is the RR ($p-1$)-form, that lives in the bulk region between the branes. 
Upon KK reduction, 
it leads to a massless 2-form $\ccc$ with  
4-d action 
\be
\label{een}
\cL_{{}_{RR}} \; = \; \ccc \wwwedge d (\cAv  + \cAh)
\; + \; 
{1\over 2 \mu^2} \; |\spc d \spc \ccc\spc |^2 \, .
\ee
This is equivalent to a St\"uckelberg mass term
for $\cAv + \cAh$.  The mass scale $\mu$ is 
 typically of order the string scale. The combination $\cAv\! +\cAh$ thus gets lifted from the low energy spectrum.  The remaining light vector boson 
\be
\label{hyper}
A_1= \cAv - \cAh 
\ee
is the hypercharge vector boson. 
This works independently of the distance
between the two branes~\cite{rrform}.
\begin{figure}
  \includegraphics[width=2.5in]{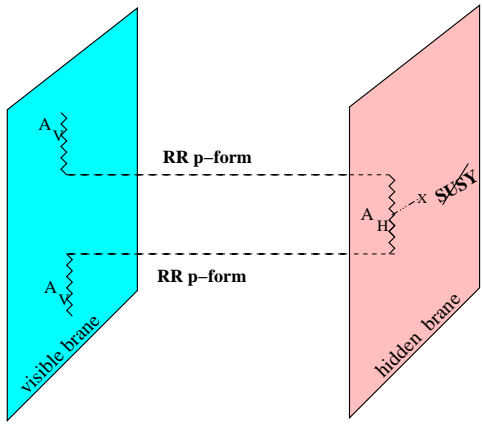} \caption{SUSY
breaking on the hidden brane is mediated to the visible sector via an
RR p-form. It
produces a mass-splitting between the $U(1)$  boson $A_V$ and
its superpartner. A more detailed account of the mechanism
is given in \cite{rrform}.  } \end{figure}

We assume that $\cAh$ is massless and that any coupling to hidden matter  
meshes with the identification of $A_1$ with the hypercharge
boson. 
This does not preclude that SUSY is broken on the hidden $U(1)$ brane.
As a concrete mechanism, consider the hidden $U(1)$ gauge kinetic term
\ba
{\cal L}_{\rm hidden}
= \int \! d^2 \theta \, \frac{1}{4}\, \fh(\varphi) \,
\cW^\alpha_{{}_H}\cW^{\ }_{{}_H,\alpha}  \; + \; c.c.
\ea
The coupling $\fh(\varphi)$ depends on closed string moduli $\varphi_m$, some of which may be in
direct contact with the region where SUSY is broken. Their F-term vevs $F_m$ induce a mass term for the superpartner
of $\cAh$,
which via the identification (\ref{hyper}),  manifests itself in the visible sector as the bino mass
\be
\label{binomass}
\tilde M_1 =  F^m \partial_m\log(\fv+ \fh).
\ee
We conclude that:
{\it  The bino mass plays a special role in phenomenological 
D-brane models with sequestered SUSY breaking.}  

\medskip

{\it UV Initial Conditions:}
The SUSY breaking F-term 
vevs $F^m$  of the closed string moduli are expressed in terms of supergravity data as $F^m =   e^{K/2} {K}^{mn} D_n W$,
where $K$ is the K\"ahler potential and $W$ the superpotential 
evaluated at the local minimum that specifies the compactification
geometry.
$K^{mn}$ is the inverse of the K\"ahler metric 
and $D_n = \partial_n - \partial_n K$. 
With sequestering,
The resulting flavor blind scenario is hypercharged anomaly mediation: 
only the bino mass receives a hidden sector \ref{binomass}
contribution while all other MSSM soft parameters are generated via
the rescaling anomaly. 
The size of the anomaly contributions is set by the gravitino mass
\be
\label{gravitino}
m_{3/2} = e^{K/2} W\, . 
\ee

At the high scale $M_*$, which for simplicity we assume to be
the GUT scale, 
we adopt the following initial conditions
for the soft  masses  and trilinear couplings
\ba
M_1 \is \tilde M_{1} \, + \, \frac{b_1 g^2  _1}{8\pi^2} \, m_{3/2} \, ;
\\
M_a \is  \frac{b_a g^2  _a}{8\pi^2} \, m_{3/2}, \qquad \mbox{\small $a
=2,3$}\, ;\\
m_i^2 \is-\frac{1}{32\pi^2} \frac{d\gamma_i}{d\log \mu}\, m_{3/2}^2 \,;
\\ A_{ijk} \is -\frac{\gamma_i + \gamma_j  + \gamma_k}{16\pi^2}\; m_{3/2}\,
. 
\ea
Here $b_a$ are the beta function coefficients, and $\gamma_i$
the anomalous dimensions of $Q_i$, evaluated at $M_{GUT}$.
Upon RG evolution, all hypercharged particles receive
mass contributions at one loop via their interaction with the $A_1$
vector multiplet. 

The relative size of the hypercharge and 
anomaly  contributions is determined by the ratio 
\be
\label{ratio}
\alpha \equiv  \tilde{M}_1 /m_{3/2}\, . 
\ee
Hypercharge mediation dominates when $\alpha$ is larger compared to
$1/4\pi$  , AMSB when $\alpha$ is very small.  
Both limits can be realized, but neither produces an acceptable spectrum. 
We will therefore assume that neither mechanism is negligible relative to the other. 
This is not an unreasonable assumption. 
Eqns. (\ref{binomass}) and (\ref{gravitino})
show that the value of  $\alpha$ is sensitive to the 
form of the superpotential $W$, moduli stabilization mechanism, and  SUSY
breaking mechanism. In the dilaton dominated limit $\alpha \lesssim
\sqrt{3}$ \
\cite{Abel:2000bj} ;
in KKLT-type scenarios, a typical value is $\alpha \sim 1/4\pi^2$
\cite{Choi:2005ge}. As we will see shortly, 
hypercharged anomaly mediation works optimally
in the intermediate range $0.05 \lesssim~|\alpha|~\lesssim~0.25$. 

\newcommand{\cC}{\mbox{\small $\phi$}}
\newcommand{\cQ}{\mbox{\small $Q$}}

\medskip

{\it RG Flow and Spectrum:} 
The free parameters are
\begin{equation}
 m_{3/2} , \;\; \alpha, \;\; \tan \beta, \;\; {\rm sign(\mu)}.
\end{equation}
Here $\tan \beta$ replaces the $B_\mu$ parameter and the magnitude of
the $\mu$ is fixed by requiring electroweak symmetry breaking (EWSB) and the measured value of the mass
of the Z boson. Thus hypercharged anomaly mediation is a highly
predictive scenario. 

\begin{figure}
\includegraphics[width=3.4in]{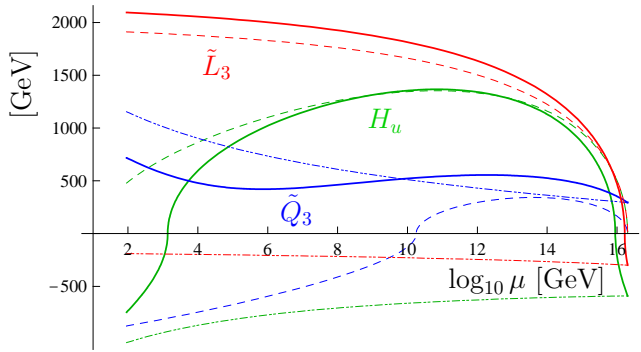}
\caption{Renormalization group running of $m_{H_u}$ (green), $m_{Q_3}$ (blue) and $m_{L_3}$ (red)
 for $\tan \beta = 10$,
$ m_{3/2} = 50$ TeV and $\alpha = 0.2$ for $M_\star = M_{GUT}$.
We define $m_{H_u} \equiv m_{H_u}^2/\sqrt{|m_{H_u}^2|} $ and similarly for $m_{Q_3}$ and $m_{L_3}$. 
The contribution of pure hypercharge mediation 
is given by dashed lines and  
the separate contribution from anomaly mediation 
is represented by the corresponding dotted lines.
}
\label{fig:RG_HuQL}
\end{figure}

Fig.~\ref{fig:RG_HuQL} depicts the separate and combined
contributions of hypercharge and anomaly mediation to the RG running
of some characteristic soft parameters --  the mass squared of the
left-handed stop, left-handed stau and  Higgs-up --
for $m_{3/2} = 50$ TeV, $\alpha = 0.2$ and $\tan \beta =10$.
In the RG evolution to the weak scale,
all scalar masses receive a contribution from the bino mass (here $Y_i$ denotes
the hypercharge)
\be
\delta m_i^2(\mu) = -\frac{3}{ 10 \pi^2}\,  g_1^2 Y_i^2 M_1^2\, \log\Bigl(\frac{\mu}{M_*}\Bigr)\, .
\ee
This positive contribution dominates at the beginning of the RG evolution. 
Once sizable scalar masses are developed, 
the negative contribution from Yukawa couplings 
becomes important and can overcome the contribution from the bino mass. 
In pure hypercharge mediation, the left-handed stop mass 
squared would be driven to negative values, because out of all scalars 
its hypercharge is the smallest
and its Yukawa coupling is the largest. All other squarks and sleptons 
remain positive. 
The wino and gluino masses receive a contribution from the bino mass at 
the two loop level. 

The anomaly induced contribution to the scalar masses is given by anomalous 
dimensions, which is negative for the mass squared of the
sleptons, and  positive for the mass squared of the squarks.
Therefore, the left handed 
stop mass is pushed above the experimental limit ($\sim 100$ GeV). This
in turn is sufficient to drive $m_{H_u}^2$ to negative values and thus 
trigger EWSB. Unless the bino contribution is negligible compared to 
the anomaly contribution, sleptons will remain sufficiently heavy in the
combined scenario. The chargino mass is above the experimental limit
provided that  $m_{3/2} \gtrsim 35$ TeV.

\begin{figure}
\includegraphics[width=3.2in]{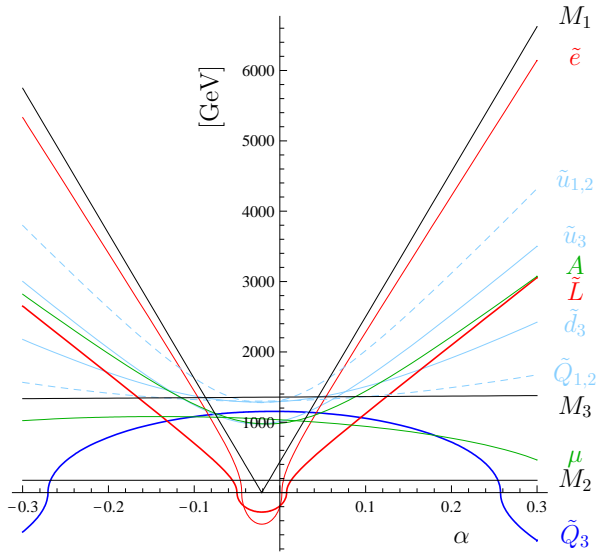}
\caption{ Plot of the spectrum of hypercharged anomaly mediation
 for $\tan \beta = 10$ and $m_{3/2} = 50$ TeV  as a function of $\alpha = M_1/m_{3/2}$. Instead of $m_{H_u}$ and $m_{H_d}$ we plot the $\mu$ term and the mass of the CP odd Higgs boson, $A$.}
\label{fig:spectrum}
\end{figure}

For $3 < \tan \beta < 50$, a  viable spectrum is obtained inside the window 
\begin{equation}
\label{window}
0.05 \lesssim  |\alpha|  \lesssim 0.25 .
\end{equation}
A region of $\alpha$ leading to a viable spectrum for $\tan \beta = 10$ 
can be read out from Fig.~\ref{fig:spectrum}  showing the spectrum as a
function of $\alpha$ for $ m_{3/2} = 50$ TeV. 
The lower bound is  given by the slepton limit and the upper bound is 
given by the limit on the stop mass. 
Hypercharge mediation dominates when 
$|\alpha|\gtrsim 0.15$.  

The mass of the light Higgs boson does not change dramatically  with
$\alpha$. For parameter choices in  Fig.~\ref{fig:spectrum} and
$|\alpha| 
\lesssim 0.2$ it
varies between 116 -- 114 GeV as
calculated by FeynHiggs2.6.2~\cite{Heinemeyer:1998yj} (with $m_t =
171$ 
GeV). It drops to 111
GeV for $\alpha \sim 0.25$ where $Q_3$
becomes very light.
Considering estimated $\pm 3$ GeV theoretical uncertainty it is consistent with the
LEP limit, 114 GeV, for $m_{3/2}$ as low as
$\sim 35$ TeV and $|\alpha| \lesssim  0.2$. Electroweak precision
tests, flavor physics observables and $g_{\mu}-2$ could impose some additional
constraints for $|\alpha| >   0.2$ and $|\alpha| < 0.05$.


The mass of the Z  boson as a result of EWSB crucially depends on the boundary condition of 
$m_{H_u}^2$ at $M_*$ and the contribution it receives from the RG evolution.  
For $\tan \beta = 10$, we have: \begin{equation} m_Z^2 \simeq -1.9
  \mu^2 - 0.0053 (\alpha -0.32) (\alpha + 0.55)
  m_{3/2}^2.  \end{equation} 
The second term is the sum of $-2m^2_{H_u}(M_*)$ and $-2\delta m_{H_u}^2$. 
As is clear from Fig. \ref{fig:RG_HuQL}, the RG contribution tends to cancel
itself.({A similar behavior
was found in models with negative stop mass squared at $M_*$
\cite{Dermisek:2006ey}.})
This is an attractive feature, not present in most other SUSY breaking scenarios, since the EWSB requires smaller amount of conspiracies among dimensionless couplings, soft SUSY breaking parameters and/or the $\mu$ term. 

We briefly comment on a few distinctive phenomenological features of
hypercharged anomaly mediation, focusing on the regime where the hypercharge
contribution dominates.
As expected, the bino is at the top of the spectrum.
Its absence from the dominant decay chains provides an obvious distinction with many other scenarios.
A second characteristic feature is the
large left-right splitting of the sfermions resulting from the
difference in their hypercharge assignments, and the related fact that, among all squarks, only 
the left-handed third generation doublet is lighter than the gluino.
Left-handed stops and sbottoms thus form important links in the gluino
decay chain. 
Top rich final state 
 can be important discovery channels, as it
typically give multiple leptons and jets. 
Disentangling these top/bottom rich final states, however, 
could be quite challenging experimentally, since 
due to large multiplicity and combinatorics, the typical top
reconstruction method is expected to  
suffer from very low efficiency. Improved 
reconstruction techniques are currently under development.
Distinguishing the left-handed stops from the right-handed stops, and
thereby uncovering the left-right 
asymmetry of the spectrum, is another non-trivial challenge. One
possible route is to measure their  
decay branching ratios into higgsino 
and wino final states.  

The lightest supersymmetric particle (LSP) is the neutral wino (except
for tiny regions of $\alpha$ where stau or stop is the LSP) which is almost
degenerate with the  lightest charged wino. Since the wino mass is highly insensitive to $\alpha$, the resulting cosmological features of our 
model, including the possibility of generating the correct dark matter density,
 are very similar to other AMSB scenarios \cite{Gherghetta:1999sw}.  


The absence of bino and sleptons, the presence of light left-handed third generation squarks,
a wino LSP, 
and potentially other observables 
combined give rise to distinctive signals at the LHC.  Finding strategies for distinguishing this
scenarios from others could still be a non-trivial challenge, however, and worth of further study.

\smallskip

{\it Discussion:}
Hypercharged anomaly mediation is a flavor blind mechanism for
communicating SUSY breaking between a geometrically
sequestered hidden and visible sector. It is a highly predictive
scenario and relies on  two known long distance forces in Nature.
In string models with the MSSM and hidden sector localized on D-branes,
the special role of the bino is geometrically well-motivated, given that 
-- via the RR-form mechanism 
\cite{rrform} -- 
only the superpartners of abelian gauge bosons can receive a mass contribution 
from the sequestered sector.

Hypercharged anomaly mediation
 predicts a low energy spectrum, that is quite insensitive to details of the 
high scale physics.
It would clearly be of interest to find concrete string models 
in which the ratio $\alpha$ between the bino and gravitino mass 
naturally ends up in the phenomenologically optimal range
(\ref{window}). We expect that such models can be constructed,
 though doing so will require a much more detailed set-up than
considered here.

Besides hypercharged anomaly mediation it is possible to extend the model by an additional $U(1)'$
which can communicate SUSY breaking to the MSSM sector, c.f.
 \cite{Langacker:2007ac}.
Among possible $U(1)'$, a combination of 
$U(1)_Y$ and $U(1)_{B-L}$ is a natural generalization. 
Furthermore, in models with a PQ-like $U(1)'$, the $\mu$ term can be generated dynamically. This
removes the problem  with large size of the corresponding $B_\mu$ term
generated by AMSB. 

Alternatively, if the hidden sector is not completely sequestered, one can use the Giudice -- Masiero mechanism
to generate the $\mu$ and $B_\mu$  terms by gravity mediation. Additional contribution to scalar masses can be generated
to remove the 
tachyonic $\tilde{Q}_3$ problem
of pure hypercharge  mediation, and also small gaugino masses can be generated. Thus a 
combination of hypercharge mediation with some contribution from 
gravity mediation 
can easily produce a viable SUSY spectrum. 

\medskip

\acknowledgments

{\it Acknowledgements:} We acknowledge useful discussions with S. Kachru, P. Langacker, G. Paz, E. Silverstein, 
T. Volansky,  and I. Yavin.  This work 
is supported by the U.S. Department of Energy, grant
DE-FG02-90ER40542
and by NSF grant PHY-0243680. 




\begin{thebibliography}{99}

\bibitem{Blumenhagen:2006ci}
 For recent reviews of string phenomenology with D-branes, see: R.~Blumenhagen, B.~Kors, D.~Lust and S.~Stieberger,
  Phys.\ Rept.\  {\bf 445}, 1 (2007).
  R.~Blumenhagen, M.~Cvetic, P.~Langacker and G.~Shiu,
  Ann.\ Rev.\ Nucl.\ Part.\ Sci.\  {\bf 55}, 71 (2005).

\bibitem{amsb}
  L.~Randall and R.~Sundrum,
  Nucl.\ Phys.\  B {\bf 557}, 79 (1999).
  G.~F.~Giudice, M.~A.~Luty, H.~Murayama and R.~Rattazzi,
  JHEP {\bf 9812}, 027 (1998).

\bibitem{Anisimov:2002az}
  A.~Anisimov, M.~Dine, M.~Graesser and S.~D.~Thomas,
  JHEP {\bf 0203}, 036 (2002).


\bibitem{Kachru:2007xp}
  S.~Kachru, L.~McAllister and R.~Sundrum,
  JHEP {\bf 0710}, 013 (2007).

\bibitem{Schmaltz:2006qs}
  M.~Schmaltz and R.~Sundrum,
  JHEP {\bf 0611}, 011 (2006).
\bibitem{Buican:2006sn}
  M.~Buican, D.~Malyshev, D.~R.~Morrison, H.~Verlinde and M.~Wijnholt,
  JHEP {\bf 0701}, 107 (2007).
\bibitem{rrform}{
H.~Verlinde, L-T.~Wang, M. Wijnholt and I.~Yavin, ``A Higher Form 
(of) Mediation'', arXiv:0711.3214 [hep-th].}

\bibitem{Abel:2000bj}
  S.~A.~Abel, B.~C.~Allanach, F.~Quevedo, L.~Ibanez and M.~Klein,
  JHEP {\bf 0012}, 026 (2000).

\bibitem{Choi:2005ge}
  K.~Choi, A.~Falkowski, H.~P.~Nilles and M.~Olechowski,
  Nucl.\ Phys.\  B {\bf 718}, 113 (2005).
\bibitem{Heinemeyer:1998yj}
  S.~Heinemeyer, W.~Hollik and G.~Weiglein,
  Comput.\ Phys.\ Commun.\  {\bf 124}, 76 (2000).
\bibitem{Dermisek:2006ey}
  R.~Dermisek and H.~D.~Kim,
  Phys.\ Rev.\ Lett.\  {\bf 96}, 211803 (2006).
\bibitem{Gherghetta:1999sw}
  T.~Gherghetta, G.~F.~Giudice and J.~D.~Wells,
  Nucl.\ Phys.\  B {\bf 559}, 27 (1999).
\bibitem{Langacker:2007ac} 
A recent phenomenological study of pure $U(1)'$ mediation is: P.~G.~Langacker, G.~Paz, L.~T.~Wang and I.~Yavin, 
arXiv:0710.1632 [hep-ph].
\end{thebibliography}
\end{document}